\documentclass[11pt]{article}
\usepackage{authblk}
\usepackage{amsfonts,amsmath,amssymb,amsthm}
\usepackage{latexsym}
\usepackage{mathrsfs}
\usepackage{tikz}
\usepackage{color}
\usepackage{multirow}

\usepackage[utf8]{inputenc}

\usepackage{amsmath} 
\usepackage{dsfont}
\usepackage[english]{babel}
\usepackage[T1]{fontenc}
\usepackage{physics}
\newtheorem{rem}{Remark}

\makeatletter
\newsavebox{\@brx}
\newcommand{\llangle}[1][]{\savebox{\@brx}{\(\m@th{#1\langle}\)}%
  \mathopen{\copy\@brx\kern-0.5\wd\@brx\usebox{\@brx}}}
\newcommand{\rrangle}[1][]{\savebox{\@brx}{\(\m@th{#1\rangle}\)}%
  \mathclose{\copy\@brx\kern-0.5\wd\@brx\usebox{\@brx}}}
\makeatother

     \numberwithin{equation}{section}

\title{\bf {The $SU(3)\supset SO(3)$ missing label problem\\ and the analytical Bethe ansatz}}
\renewcommand*{\Affilfont}{\normalsize\small}
\author[1]{Nicolas Crampé\,}
\author[2]{Dounia Shaaban Kabakibo\,}
\author[3]{Luc Vinet\,\vspace{.5em}}
\affil[1]{Institut Denis-Poisson CNRS/UMR 7013 - Université de Tours -
Université
d'Orléans, \newline\vspace{.9em}
Parc de Grandmont, 37200 Tours, France.}
\affil[2,3]{Centre de Recherches Math\'ematiques, Universit\'e de Montr\'eal,
\newline\vspace{.9em}
P.O. Box 6128, Centre-ville Station, Montr\'eal (Qu\'ebec), H3C 3J7, Canada.}
\affil[3]{Insitut de valorisation des donn\'ees (IVADO), Montr\'eal
(Qu\'ebec), H2S 3H1, Canada. \newline\vspace{.9em}}

{
\makeatletter
\renewcommand\AB@affilsepx{: \protect\Affilfont}
\makeatother
\affil[ ]{E-mail addresses}
\makeatletter
\renewcommand\AB@affilsepx{, \protect\Affilfont}
\makeatother
\affil[1]{crampe1977@gmail.com}
\affil[2]{dounia.shaaban.kabakibo@umontreal.ca}
\affil[3]{vinet@crm.umontreal.ca}
}

\begin{document}
\maketitle

\vspace{15mm}

\abstract{}
The missing label for basis vectors of $SU(3)$ representations corresponding to the reduction $SU(3) \supset SO(3)$ can be provided by the eigenvalues of $SO(3)$ scalars in the enveloping algebra of $su(3)$. There are only two such independent elements of degree three and four. It is shown how the one of degree four can be diagonalized using the analytical Bethe ansatz.
\vspace{2cm}
\section{Introduction}
This paper bears on the missing label problem for the basis states of $SU(3)$ representations corresponding to the reduction $SU(3) \supset SO(3) \supset SO(2)$. This is a question that has been studied for a long time and a topic on which there is a vast literature. The reader may consult \cite{MPSW} for a review of where the subject stood roughly 50 years ago and for finding many references. The problem has not lost its interest through the years in view of its practical importance and has kept being explored. Among more recent reports we may cite \cite{Tolstoy, Campoamor, PYLD, VBGDGPK} 
where additional references will be found and, in particular \cite{JZ}, which connects the missing label problem to quantum groups and integrable models and will prove of relevance in the following.

In a nutshell, the problem has to do with the fact that generally, the irreducible representations (irreps) of $SO(3)$ are not multiplicity free in the irreps of $SU(3)$. Hence the values of the two Casimir elements of $SU(3)$, and the quantum numbers that label the basis vectors of $SO(3)$ irreps do not suffice to characterize the basis vectors of $SU(3)$ irreps corresponding to the reduction $SU(3) \supset SO(3) \supset SO(2)$. In cases where the same $SO(3)$ irrep occurs more that once, clearly an additional multiplicity index is needed.

Of the various resolutions of the problem that have been proposed, the one we will concentrate on has the missing label provided by the eigenvalues of an operator belonging to the centralizer of $so(3)$ in the enveloping algebra $U(su(3))$ of the Lie algebra of $SU(3)$. It has been shown \cite{JMPW} that the integrity basis for this centralizer consists of two elements $x$ and $y$ of degree three and four respectively in the generators of $su(3)$. This approach has the merit of yielding an orthogonal set of basis vectors at the expense of having labels -the eigenvalues- that are not integer and somewhat difficult to compute. The operator $x$ was first introduced in \cite{BM}, while $y$ can be traced back to lecture notes by Racah \cite{R}. We shall also make use of the resolution provided by Bargmann and Moshinsky who constructed an $SU(3) \supset SO(3)$ basis bearing their names in terms of elementary permissible diagrams. In this BM basis, the additional label is integer but the vectors are not orthogonal.

Our attention will be focused on the spectrum of $y$. Numerical results were obtained in \cite{JMPW} by relying on the Gelf'and--Tseitlin states. Another method based on shift operators is described in \cite{H, DVV}. We shall here add to this venerable topic by showing how the fourth order labelling operator $y$ can be diagonalized by calling upon the analytical Bethe ansatz (BA).

Recently, the Bethe ansatz and its variants, which are cornerstones in the study of  (quantum) integrable models, have found their way in representation theory. They have in particular appeared in two situations that bear a very close relation to the problem of finding the spectrum of an $SO(3)$ scalar in $U(su(3))$: i. the diagonalization in \cite{BCSV} using the algebraic Bethe ansatz of the Lie--Heun operator introduced in \cite{PW_Heun} as labeling operator of a non-subgroup basis of $O(3)$. ii. the determination in \cite{CPV2} with the nested Bethe ansatz of the eigenvalues of an operator providing the missing label in the $SU(3)$ Clebsch--Gordan problem.

With respect to the diagonalization of $y$ using Bethe ansatz techniques, as already alluded to, a useful observation is the one made in \cite{JZ} where the operator $y$ was found to lie in a Bethe subalgebra of the twisted Yangian $Y(gl(3),so(3))$. Given this identification, we shall show that it is possible to diagonalize $y$ using the analytical Bethe ansatz. This method has been developed previously to find the spectrum of different quantum integrable spin chains (see for example \cite {Reshetikhin1,Reshetikhin2,Reshetikhin3,VR,MN, AMN}). It lays on assuming a particular form of the eigenvalues. The Bethe equations are obtained by imposing their analyticity, which explains the name of the method. We are going in fact to generalise this method in order to diagonalize the whole Bethe subalgebra. The hypothesis we make for the form of the eigenvalue is inspired from work done in \cite{MTV} where the Bethe subalgebra of the Yangian has been diagonalized using the nested algebraic Bethe ansatz.

The paper will unfold as follows. Section 2 will introduce the missing label operators $x$ and $y$ of third and fourth order and recall the remarkable relations found by Lehrer and Racah that they obey. It is worth stressing that strikingly, the algebra thus defined has also appeared recently \cite{CPV} as the centralizer of the diagonal embedding of $U(sl(3))$ in $U(sl(3))^{\otimes 2}$. Section 3 will look at the actions of $x$ and $y$ in the BM basis which will be found to be tridiagonal. Complementing the results of Bargmann and Moshinsky who had obtained the matrix elements of $x$, we provide those of $y$. Section 4 will present the extension of the analytical Bethe ansatz that is required to diagonalize $y$. The results of section 3 will be used to offer a comparison between some of the eigenvalues computed with the Bethe ansatz approach and the corresponding ones obtained from the direct diagonalization of the tridiagonal matrix expressing $y$ in the BM basis. Throughout the paper, we shall often interchangeably refer to Lie groups and their algebras denoting the former by capital letters and  the latter by lower case ones.

\section{The $SU(3) \supset SO(3)$ missing label operators and their algebra}
In this section, we establish notations and introduce the missing label operators $x$ and $y$ in $U(su(3))$. We then present the double commutator relations that these operators verify. 

The complexified Lie algebra $su(3)$ is generated by  $E_{ij}$ ($1 \leq i, j \leq 3$) satisfying the defining relations
\begin{equation}
[E_{ij},E_{kl}]=\delta_{jk}E_{il}-\delta_{il}E_{kj},
\end{equation}
with the extra condition 
\begin{equation}
\sum_{i=1}^3{E_{ii}}=0.
\end{equation}

When considering the subgroup chain $SU(3) \supset SO(3)$, it is convenient to use a basis of $su(3)$ \cite{R} generated by the five independent components $T_{mn}$ ($1\leq m, n \leq 3$) of the  symmetrical traceless tensor $\mathbf{T}$  and the three components $L_k$ ($1\leq k \leq3$) of the angular momentum vector $\mathbf{L}$ generating the $so(3)$ subalgebra. They satisfy the following commutation rules:
\begin{align}
\nonumber  [L_k,L_n]&=i\epsilon_{knp}L_p, \\
\nonumber  [L_k,T_{mn}]&=i(\epsilon_{kmp}T_{pn}+\epsilon_{knp}T_{mp}),\\
\nonumber  [T_{mn},T_{pq}]&=i(\delta_{mp}\epsilon_{nqr}+\delta_{mq}\epsilon_{npr}+\delta_{np}\epsilon_{mqr}+\delta_{nq}\epsilon_{mpr})L_r,
\end{align}
where $\epsilon_{kmn}$ is the Levi-Civita symbol and the sums over the repeated indices are understood.
The previous generators can be written in terms of  $E_{ij}$. Indeed, the $so(3)$ generators are written as follows
\begin{align}
 &L_{1} = \frac{E_{12} - E_{32} + E_{21} - E_{23}}{\sqrt{2}},& 
 &L_{2} = -\frac{i (E_{12} + E_{32} - E_{21} - E_{23})}{\sqrt{2}},& &L_3 = E_{11} - E_{33},
\end{align}
and the five independent components of $\mathbf{T}$ are 
\begin{align}  
\nonumber 
&T_{11} = E_{11} + E_{33} +  E_{13} +  E_{31} ,& 
 &T_{22} = E_{11} + E_{33}-  E_{13} -  E_{31} ,& &T_{12} = i(E_{31}-E_{13} ),\\
\nonumber   &T_{13} =- \frac{E_{12} + E_{32} + E_{21} + E_{23}}{\sqrt{2}},&
 &T_{23} = \frac{i (E_{12} - E_{32} - E_{21} + E_{23})}{\sqrt{2}}.& & 
\end{align}
The two Casimir elements of $su(3)$ are
\begin{eqnarray}
&& g_2=L_kL_k+\frac{1}{2}T_{mn}T_{nm}, \qquad g_3=\frac{1}{3}T_{mn}T_{np}T_{pm}-L_mT_{mn}L_n.
\end{eqnarray}
Their values characterize the finite irreps of $SU(3)$. For the chain $SU(3) \supset SO(3)$, one also considers the Casimir element of $so(3)$:
\begin{eqnarray}
 && \mathbf{L}^2=L_kL_k.
\end{eqnarray}
 A supplementary operator is needed in order to distinguish the different copies of the $SO(3)$ irrep within a given $SU(3)$ irrep.  The degeneracy can be lifted using one of the two following operators belonging to the universal enveloping algebra of $su(3)$\cite{BM, R, JMPW}:
\begin{eqnarray}
 && \bar{x}=L_mT_{mn}L_n, \qquad \bar{y}=L_mT_{mn}T_{np}L_p.
 \end{eqnarray}
One generator of $SO(3)$, the angular momentum projection $L_3$, completes the characterization of the basis vectors.

For reasons of simplicity, we consider the following shifted operators:
\begin{eqnarray}
&&x=\frac{1}{16}(\bar{x}+\frac{g_3}{4}), \qquad y=-\frac{1}{64}\Big (\bar{y}+3-g_2-\frac{\mathbf{L}^2}{3}(9+g_2-\mathbf{L}^2)\Big).
\end{eqnarray}
Then by using the defining relations of $su(3)$, one can show that $x$ and $y$ satisfy the following relations:
\begin{align} 
& \nonumber [x,[x,y]]=-6y^2+a_2x^2+a_5x+a_8,\\
& [y,[y,x]]=2x^3+a_2\{x,y\}+a_5y-a_6x-a_9,
\label{eq:algebra}
\end{align}
where the coefficients are given in terms of $g_2, g_3, \mathbf{L}^2$ as
\begin{align}
\nonumber a_2&=\frac{g_2+2\mathbf{L}^2-18}{8},\\
\nonumber a_5&=\frac{(4\mathbf{L}^2-g_2)g_3}{256},\\
\nonumber a_6&=\frac{1}{6144}\left(8g_2(3-\mathbf{L}^2)(2g_2-\mathbf{L}^2-21)+9 g_3^2-16 \mathbf{L}^2 ((\mathbf{L}^2-18) \mathbf{L}^2+63)+1296\right),\\
\nonumber a_8&=\frac{1}{98304}\big(
16\left(\mathbf{L}^2-3\right)^2\left(g_2^2+\left(\mathbf{L}^2-9\right)\left(\mathbf{L}^2-3\right)\right)-16\left(2\mathbf{L}^2-3\right)\left(\left(\mathbf{L}^2\right)^2-6\mathbf{L}^2-18\right)g_2\\
\nonumber&+3g_3^2(18-10\mathbf{L}^2+g_2)+2592(\mathbf{L}^2-1)\big),\\
\nonumber a_9&=\frac{1}{196608} \left(4g_2 \left(\mathbf{L}^2-3\right)\left(2g_2-7 \mathbf{L}^2-3\right)
-3 g_3^2-16 \mathbf{L}^2 \left(\left(\mathbf{L}^2-18\right) \mathbf{L}^2+9\right)\right)g_3.
\end{align}
These relations have been previously given in \cite{R} without explicit expressions for the coefficients $a_i$. Let us note that a more general version of this algebra also appears in \cite{CPV}. In the latter paper, a central element $\Omega$ has been discovered. By specializing this result to our case, one gets that
\begin{equation}
\Omega =b_1x+b_2y+b_3x^2+b_4\{x,y\}
+b_5 y^2+b_7 xyx-x^4+4y^3+[x,y]^2
\end{equation}
with
\begin{equation}
 b_1= 6 a_5  + 2 a_9\,,\ \ \ b_2= -2a_6 - 2 a_8\,,\ \ \ b_3= 6 a_2 + a_6\,,\ \ \ b_4=-a_5\,,\ \ \ b_5= 8 a_2 -24\,,\ \ \ b_7= -2 a_2 +12\ ,
\end{equation}
commutes with $x$ and $y$. It also commutes with $g_2, g_3, \mathbf{L}^2, L_3$ and is proportional to the identity in a given irrep.
This is indeed compatible with equation (139c) of \cite{R} stating that the square of the commutator $[x,y]$ is a function of $x, y, \mathbf{L}^2, g_2, g_3$.

\section{Differential realization of the missing label operators $x$ and $y$}
In this section, we review a well-known polynomial basis for the irreducible representations of $SU(3) \supset SO(3)$, namely the Bargmann--Moshinsky basis. We present expressions for the matrix elements of $x$, $y$ in this basis.
\subsection{Irreducible representations of $su(3)$}

A finite dimensional irreducible representation of $su(3)$ is labeled by a pair of  non-negative integers $(\lambda, \mu)$. These give the Cartan labels of the highest weight vector $V_{\lambda,\mu}$ satisfying:
\begin{align}
   \label{hw1} E_{ii}V_{\lambda,\mu}&=\alpha_{ii}V_{\lambda,\mu} \qquad \text{for } 1\leq i\leq3,\\
     \label{hw3} E_{ij}V_{\lambda,\mu}&=0  \qquad \text{for } 1\leq i< j\leq3,
\end{align}
with
\begin{equation} \label{alphaii}
\alpha_{11}=\frac{\mu+2\lambda}{3}, \quad \alpha_{22}=\frac{\mu-\lambda}{3}, \quad \alpha_{33}=-\alpha_{11}-\alpha_{22}=-\frac{2\mu+\lambda}{3}. 
\end{equation}
This representation correspond to a Young diagram with $\lambda+\mu$ boxes in the first row and $\mu$ boxes in the second row. 
The dimension of these irreps is 
\begin{equation}
    d_{(\lambda,\mu)}=\frac{1}{2}(\mu+1)(\lambda+1)(\lambda+\mu+2).
\end{equation}
For a detailed construction of $su(3)$ irreps using the highest weight vector, see for example \cite{FH}.

The two Casimir operators of $su(3)$ become proportionnal to the identity in this irrep and are related to $\lambda, \mu$ by:
\begin{align} \label{g2andg3}
& g_2=\frac{4}{3}(\lambda^2+\mu^2+\lambda\mu+3\lambda+3\mu),
& g_3=\frac{8}{27}(\lambda-\mu)(3+\lambda+2\mu)(3+2\lambda+\mu).
\end{align}
A given $so(3)$ representation characterized by the integer $L$ (such that $\mathbf{L}^2=L(L+1)$), appears $d_{(\lambda,\mu)}^L$ times in the $su(3)$ irrep.  This degeneracy has been computed in $\cite{R2}$ and is given by:
\begin{equation}
d_{(\lambda,\mu)}^L=\left[\frac{\lambda+\mu-L+2}{2}\right]_{\geq 0}-\left[\frac{\lambda-L+1}{2}\right]_{\geq 0}-\left[\frac{\mu-L+1}{2}\right]_{\geq 0},
\end{equation}
where $[\dots]_{\geq 0}$ is zero for any negative number and gives the positive integer part otherwise.
One can then check that we indeed have:
\begin{equation}
    d_{(\lambda,\mu)}=\sum_L (2L+1) d_{(\lambda,\mu)}^L .
\end{equation}
\subsection{The Bargmann--Moshinsky basis}
The Bargmann--Moshinsky (BM) basis \cite{BM} consists in a set of states for the general irrep ($\lambda$,$\mu$) of $SU(3) \supset SO(3)$ reduction which depends on the supplementary label $\alpha$ that distinguishes the different copies of $SO(3)$ irreps. The highest states of all $SO(3)$ multiplets in this basis are given by
\begin{equation}
\left | 
\begin{matrix}
\lambda &\mu \\
\alpha & L& L
\end{matrix}
\right \rangle_B=
\begin{cases}\displaystyle
\eta_1^{\lambda-2\alpha}\;\eta_{12}^{L-\lambda+2\alpha}\;s_1^{\alpha}\;s_{12}^{\frac{1}{2}(\mu-L+\lambda-2\alpha)}, \; &\lambda+\mu-L \text{      even}\vspace{3mm} \\

\eta_1^{\lambda-2\alpha-1}\;\eta_{12}^{L-\lambda+2\alpha}\;w\;s_1^{\alpha}\;s_{12}^{\frac{1}{2}(\mu-L+\lambda-2\alpha-1)}, \; &\lambda+\mu-L \text{  odd}
\end{cases}
\end{equation}
where:
\begin{align}
\nonumber  &\eta_{12}=\eta_1\zeta_2-\zeta_1\eta_2,&  &\zeta_{12}=\eta_1\xi_2-\xi_1\eta_2,& &\xi_{12}=\zeta_1\xi_2-\xi_1\zeta_2,\\
  &s_1=\zeta_1^2-2\eta_1\xi_1,&  &s_{12}=\zeta_{12}^2-2\eta_{12}\xi_{12},& &w=\eta_1\zeta_{12}-\zeta_1\eta_{12}.
\end{align}
In this realization, the generators $E_{ij}$ of the $su(3)$ algebra are given by
\begin{equation}
E_{ij}=v_i\frac{\partial}{v_j}+w_i\frac{\partial}{w_j}-\frac{\delta_{ij}}{3}\left(\lambda+2\mu\right)
\end{equation}
with
\begin{eqnarray}
\qquad v=(-\eta_1,\zeta_1,\xi_1),\qquad w=(-\eta_2,\zeta_2,\xi_2).
\end{eqnarray}
The rest of the basis elements 
\begin{eqnarray}
\qquad \left | 
\begin{matrix}
\lambda &\mu \\
\alpha & L& m
\end{matrix}
\right \rangle_B \qquad \text{where} \qquad L_3 \left | 
\begin{matrix}
\lambda &\mu \\
\alpha & L& m
\end{matrix}
\right \rangle_B =m\left |\begin{matrix}
\lambda &\mu \\
\alpha & L& M
\end{matrix}
\right \rangle_B,
\end{eqnarray}
 are obtained by repeatedly acting with $L_{-}=\frac{iL_2-L_1}{\sqrt{2}}
$.
The highest weight state $V_{\lambda,\mu}$ in a given  $su(3)$ irrep is obtained by setting $m=L=\lambda+\mu$ and $\alpha =0$ in the BM vectors.
The label $\alpha$ is a positive integer and is chosen such as all exponents are positive. It thus lays in the range
\begin{equation}\label{alpharange}
\text{max} \Big(0,\frac{\lambda-L}{2}+\frac{1-(-1)^{\lambda-L}}{4} \Big)\leq \alpha \leq \text{min} \Big(\frac{\lambda-1}{2}+\frac{(-1)^{\mu+L}}{4}((-1)^\lambda+1),\frac{\lambda+\mu-L}{2}+\frac{(-1)^{\lambda+\mu+L}-1}{4} \Big).
\end{equation}

\begin{rem}
The degeneracy $ d_{(\lambda,\mu)}^L$ can also be found using equation \eqref{alpharange}:
\begin{align}
\nonumber d_{(\lambda,\mu)}^L=&\min\left[\frac{L}{2}+\frac{1}{4}+\frac{(-1)^L}{4}\left((-1)^\lambda+(-1)^\mu+(-1)^{\lambda+\mu}\right),\frac{\mu}{2}+\frac{1}{2}+\frac{(-1)^{\lambda+L}}{4}(1+(-1)^\mu )\right.,\\
&\left.\frac{\lambda+1}{2}+\frac{(-1)^{\mu+L}}{4}((-1)^\lambda+1),\frac{\lambda+\mu-L}{2}+\frac{(-1)^{\lambda+\mu+L}+3}{4}\right].
\end{align}
The group of linear transformations acting on the three parameters  $\lambda, \mu, L $ and preserving the size of the missing label matrices is the group of symmetry of the square. It is generated by the two permutations $r=(1,4)$ and $s=(13)(24)$ of the four possible values of $d_{(\lambda,\mu)}^L$ and are explicitly given by:

\begin{align}
   \nonumber \text{r : }L&\rightarrow \lambda+\mu-L+1-\frac{(-1)^{\lambda+L}+(-1)^{\mu+L}}{2} &&\; & \text{s : }L&\rightarrow \lambda+\frac{1-(-1)^{\lambda+L}}{2}\\
\nonumber \lambda &\rightarrow \lambda &&\; ,  & \lambda &\rightarrow L-\frac{1-(-1)^{\lambda+L}}{2}\\
 \mu &\rightarrow \mu &&\; &  \mu &\rightarrow \lambda+\mu -L +\frac{1-(-1)^{\lambda+L}}{2}.  
\end{align}
\end{rem}

\subsection{The operators $x$ and $y$ in the Bargmann--Moshinsky basis}As mentioned earlier, it is not possible to construct a basis for $SU(3)\supset SO(3)$ such that the operators $x$ and $y$ have analytical eigenfunctions \cite{R}. Nevertheless, they both take the form of tridiagonal matrices in the BM basis.
The resulting matrices can then be diagonalized numerically to find the spectrum.

The two non trivial expansions are: 
\begin{equation}
\qquad x\left | 
\begin{matrix}
\lambda &\mu \\
\alpha & L& m
\end{matrix}
\right \rangle_B=\sum_{\alpha'}\left | 
\begin{matrix}
\lambda &\mu \\
\alpha' & L& m
\end{matrix}
\right \rangle_B\beta_{\alpha'\alpha},
\qquad y\left | 
\begin{matrix}
\lambda &\mu \\
\alpha & L& m
\end{matrix}
\right \rangle_B=\sum_{\alpha'}\left | 
\begin{matrix}
\lambda &\mu \\
\alpha' & L& m
\end{matrix}
\right \rangle_B\gamma_{\alpha'\alpha}.
\end{equation}
The non vanishing coefficients $\beta_{\alpha'\alpha}$ have been computed in \cite{BM}. In our notation, they read as:

$\lambda+\mu+L$ even:
\begin{align}
\beta_{\alpha+1,\alpha}^{(e)}&=\frac{1}{8}(2\alpha-\lambda)(2\alpha-\lambda+1)(2\alpha-\mu+L-\lambda),\\
\beta_{\alpha-1,\alpha}^{(e)}&=\frac{1}{4}\alpha(2\alpha+L-\lambda)(2\alpha+L-\lambda-1),\\
\nonumber \beta_{\alpha,\alpha}^{(e)}&=\frac{1}{216} \Big(27 (2 \alpha-\lambda+L) \left(8 \alpha^2-2 \alpha (3\lambda-L+\mu)+\lambda^2+\mu (\lambda+L)+L+1\right)\\
& -9L(L+1) \left(\lambda+2 \mu+\frac{3}{2}\right)+(\lambda-\mu) (2 \lambda+\mu+3) (\lambda+2 \mu+3)+27 L \mu\Big),
\end{align}

$\lambda+\mu+L$ odd:
\begin{align}
\beta_{\alpha+1,\alpha}^{(o)}&=\frac{1}{8}(2\alpha-\lambda+1)(2\alpha-\lambda+2)(2\alpha-\mu+L-\lambda+1),\\
\beta_{\alpha-1,\alpha}^{(o)}&=\frac{1}{4}\alpha(2\alpha+L-\lambda)(2\alpha+L-\lambda-1),\\
\nonumber \beta_{\alpha,\alpha}^{(o)}&=\frac{1}{216} \Big(27 (2 \alpha-\lambda+L) \left(8 \alpha^2-2 \alpha (3 \lambda-L+\mu-3)+\lambda^2+\lambda (\mu-1)+L \mu+2 L+5\right)\\
&-9 L (L+1) \left(\lambda+2 \mu+\frac{3}{2}\right)+(\lambda-\mu) (2 \lambda+\mu+3) (\lambda+2 \mu+3)+27 \lambda+27 L (\mu-1)+54\Big),
\end{align}

We furthermore carried on the same computation for the operator $y$ and obtained the following analytical expressions for its non vanishing elements:

$\lambda+\mu+L$ even:
\begin{align} \label{ymatrixcoeffs1}
    \gamma_{\alpha+1,\alpha}^{(e)}&=\frac{1}{6}( 6 \alpha- \lambda +\mu +3 L+6)\beta_{\alpha+1,\alpha} ^{(e)},\\
\gamma_{\alpha-1,\alpha}^{(e)}&= \frac{1}{3}( 3 \alpha- 2\lambda - \mu-3)\beta_{\alpha-1,\alpha}^{(e)},\\
\nonumber \gamma_{\alpha,\alpha}^{(e)}&=\frac{1}{192} \Big(384 \alpha^4-128 \alpha^3 (5 \lambda-3 L+\mu)\\
\nonumber&+16 \alpha^2 \left(23 \lambda^2+\lambda (8 \mu-30 L-6)+6 L^2-3 L (2 \mu+1)-\mu^2-6 \mu+3\right)\\
\nonumber&-8 \alpha \big(10 \lambda^3-2 \lambda^2 (11 L-2 \mu+6)+2 \lambda \left(5 L^2+L (2-4 \mu)-\mu^2-6 \mu+2\right)+L^2 (2 \mu+3)\\
 \nonumber &+L (2 \mu-3)+2 \mu\big)+4 \lambda^2 \left(3 L^2+L (7-2 \mu)-\mu^2-6 \mu+2\right)-8 \lambda^3 (2 L+3)+4 \lambda^4\\
 &+4 \lambda \left(L^2 (\mu-1)+L (3 \mu-1)+3 (\mu+1)\right)-L^4-2 L^3+5 L^2+6 L+4 \mu^2+12 \mu-9\Big),
\end{align}
$\lambda+\mu+L$ odd:
\begin{align}
\gamma_{\alpha+1,\alpha}^{(o)}&=\frac{1}{6}(6\alpha-\lambda+\mu+3L+9)\beta_{\alpha+1,\alpha}^{(o)},\\
\gamma_{\alpha-1,\alpha}^{(o)}&=\frac{1}{6}(6\alpha-4\lambda-2 \mu-6) \beta_{\alpha-1,\alpha} ^{(o)}, \\
\nonumber \gamma_{\alpha,\alpha}^{(o)}&=\frac{1}{192} \Big(384 \alpha^4-128 \alpha^3 (5 \lambda-3 L+\mu-3)\\
\nonumber&+16 \alpha^2 \left(23 \lambda^2+\lambda (-30 L+8 \mu-36)+6 L^2+L (15-6 \mu)-\mu^2-12 \mu+12\right)\\
\nonumber &-8 \alpha \big(\lambda^2 (-22 L+4 \mu-34)+10 \lambda^3+2 \lambda \left(5 L^2-4 L (\mu-4)-\mu^2-10 \mu+13\right)+L^2 (2 \mu-3)\\
\nonumber &+L (10 \mu-9)+10 \mu-6\big)+4 \lambda^2 \left(3 L^2+L (15-2 \mu)-\mu^2-8 \mu+14\right)-8 \lambda^3 (2 L+5)+4 \lambda^4\\ \label{ymatrixcoeffs2}
&+4 \lambda \left(L^2 (\mu-5)+L (7 \mu-13)+9 \mu-5\right)-8 L^2 \mu-L^4-2 L^3+5 L^2-24 L \mu+6 L+4 \mu^2-4 \mu-9\Big).
\end{align}
\begin{rem} With an appropriate identification of parameters,
it can be shown that the matrix elements of $x$ and $y$ have the same form as those of the missing label operators in the two fold tensor product of $su(3)$ found in $\cite{CPV2}$.
\end{rem}

\section{Analytical Bethe ansatz}
In this section, the definitions of the Yangian of $gl(3)$ and of the twisted Yangian  $Y (gl(3); so(3))$ are recalled (see e.g. \cite{MNO}). The connection, discovered in $\cite{JZ}$, between  an abelian subalgebra of $Y (gl(3); so(3))$, called Bethe subalgebra, and the missing label operator $y$ is reviewed. Exploiting this result, the spectrum of $y$ is found using the analytical Bethe ansatz. This is achieved by generalising the method used in \cite{ACDFR, ACDFR2} to diagonalize the transfer matrix associated with open spin chains with soliton non preserving boundary conditions. 
\subsection{Yangian $Y(gl(3))$, Twisted Yangian $Y (gl(3); so(3))$ and a Bethe subalgebra}
The Yangian $Y(gl(3))$ is defined \cite{Drinfeld1, Drinfeld2} as the complex associative unital
algebra generated by the generators $\{t_{ij}^{(n)}|1\leq i,j\leq 3,
n\in \mathbb{Z}_{\geq 0}\}$
 subject to the defining relations:
\begin{eqnarray}
\label{relcom}
[t_{ij}^{(r+1)}\,,\,t_{kl}^{(s)}]-[t_{ij}^{(r)}\,,\,t_{kl}^{(s+1)}]
=t_{kj}^{(r)}\,t_{il}^{(s)}-t_{kj}^{(s)}\,t_{il}^{(r)},
\end{eqnarray}
where $r,s\in \mathbb{Z}_{\geq 0}$ and $t_{ij}^{(0)}=\delta_{ij}$.
In order to encompass all the defining relations of the Yangian, it is known that we can use the FRT presentation \cite{FRT}.
It is based on the following $R$-matrix
\cite{yang, baxter1, baxter2, baxter3}
\begin{eqnarray}
  R(u) =u\;\mathbb{I}_3 \otimes \mathbb{I}_3 -
\mathbb{P}, \label{r}
\end{eqnarray}
where $\mathbb{I}_3$ is the $3\times 3$ identity matrix, $\mathbb{P}$ is the permutation operator
\begin{equation}
  \label{eq:P12}
  \mathbb{P} = \sum_{i,j=1}^3 e_{ij} \otimes e_{ji},
\end{equation}
and $e_{ij}$ is a $3\times 3$ matrix with zeros everywhere and 1 at the entry $(i,j)$.
It provides a solution to the Yang--Baxter equation:
\begin{eqnarray}
  R_{12}(u-v)\ R_{13}(u)\ R_{23}(v)
  =R_{23}(v)\ R_{13}(u)\
  R_{12}(u-v),
  \label{YBE}
\end{eqnarray}
 with $R_{12}=R\otimes\mathbb{I}_3$, $R_{23}=\mathbb{I}_3\otimes R$ and so on.
 We then define the following T-matrix:
 \begin{equation}
 T(u)=\sum_{i,j=1}^3 e_{ij}\otimes t_{ij}(u) ,
 \end{equation}
 with $t_{ij}(u)$ the following formal series in $1/u$:
 \begin{equation}
     t_{ij}(u)=\delta_{ij}+\sum_{k=1} ^ \infty \frac{t_{ij}^{(k)}}{u^k}.
 \end{equation}
 The commutation relations \eqref{relcom} can then by compactly written in the form of an FRT equation \cite{FRT}
 \begin{equation}
    R(u-v)T_1(u)T_2(v)=T_2(v)T_1(u)R(u-v).
 \end{equation}

 We now introduce the following generalised transposition $t$, related to the usual transposition $T$ by:
 \begin{equation}
     A^t=V^{-1}A^TV , \quad \text{where}\quad V=\text{antidiag}(1, 1,\dots,1).
 \end{equation}
 The $so(3)$ twisted Yangian is a subalgebra of $Y(gl(3))$, noted as $Y (gl(3); so(3))$, and is generated by elements $s_{ij}^{(k)}$ put in the following formal series
 \begin{equation}
 S(u)=\sum_{i,j}^3\sum_{k=0}^\infty  e_{ij}\otimes \frac{s_{ij}^{(k)}}{u^k}=T(u)T^t(-u).
 \end{equation}
The commutation relations of the twisted Yangian can be compactly encoded as:
 \begin{equation}
   R(u-v)S_1(u)\widetilde{R}(-u-v)S_2(v)=S_2(v)\widetilde{R}(-u-v)S_1(u) R(u-v),
 \end{equation}
 where 
 
 \begin{equation}
     \widetilde{R}(u)=R^{t_1}(u),
 \end{equation}
 and $t_1$ is the generalised transposition taken in the first space only.
 
 One can then construct an Abelian subalgebra of  $Y (gl(3); so(3))$, also called a Bethe subalgebra \cite{NO}, generated by the coefficients in $u^{-1}, u^{-2}, \dots$ of the following 3 elements $A_1(u), A_2(u), A_3(u)$:
 \begin{align}
 A_1(u)&=tr_1[S_1(u)],\label{Bethesubalgebra1}\\
 A_2(u)&=tr_{12}[(\mathbb{A}_{12}\; S_1(u)\widetilde{R}_{12}(-2u+1)S_2(u-1)],\label{Bethesubalgebra2}\\
\nonumber A_3(u)&=tr_{123}[\mathbb{A}_{123} \; S_1(u)\widetilde{R}_{12}(-2u+1)\widetilde{R}_{13}(-2u+2)\\
  & \qquad \ \times S_2(u-1)\widetilde{R}_{23}(-2u+3)S_3(u-2)],\label{Bethesubalgebra3}
 \end{align}
 where the $\mathbb{A}$s are antisymmetrization operators:
 \begin{align}
     \mathbb{A}_{12}&=\mathbb{I}-\mathbb{P}_{12}\\
     \mathbb{A}_{123}&=\mathbb{I}-\mathbb{P}_{12}-\mathbb{P}_{13}-\mathbb{P}_{23}+\mathbb{P}_{12}\mathbb{P}_{23}+\mathbb{P}_{13}\mathbb{P}_{23}
 \end{align} 
 and $tr_1[\dots]$ is the trace taken over the first space, $tr_{12}[\dots]$  is the trace taken over the first and second spaces, and so on.
  The three series $A_i$ satisfy the following commutation relations \cite{NO}
 \begin{equation}
     [A_i(u),A_j(v)]=0 \qquad \text{for} \qquad i,j=1,\dots 3,
 \end{equation}
 which proves that the coefficients of the series are mutually commutative and form an Abelian subalgebra. 
 \subsection{Evaluation representation of $Y(gl(3))$}
 The construction of representations of the Yangians is based on the following algebra homomorphism from $Y(gl(3))$ to the universal enveloping algebra of $gl(3)$ 
 \begin{equation}
     T_{ij}\longmapsto \delta_{ij}+\frac{E_{ij}}{u}.
 \end{equation}
It is called the evaluation representations of $Y(gl(3))$.
In this realisation, the element $A_1(u)$ becomes
\begin{equation}\label{A1Eij}
    A_1(u)=3 - \frac{  2E_{32}E_{12} + 2E_{31}E_{13} + 2E_{21}E_{23}+2E_{33}E_{11} + E_{22}^2 + E_{11} - E_{33}}{u^2}
\end{equation}
The same computation can be done for $A_2(u)$ and $A_3(u)$.
Then the elements \eqref{Bethesubalgebra1}-\eqref{Bethesubalgebra3} can be rewritten as follows \cite{JZ}: 
\begin{align}
 A_1(u)&=3-\frac{g_2-2\mathbf{L}^2}{2u^2}, \label{A1su3}\\
 A_2(u)&=(1 - 2u)\left(\frac{32y}{( u-1)^2u^2} + 
   6 + \frac{(\mathbf{L}^2 - g_2)(12u^2 + \mathbf{L}^2 - 12u)}{6( u-1)^2u^2} + \frac{g_2^2 - 
      8\mathbf{L}^2 - 4g_2 + 12}{8( u-1)^2u^2}\right), \label{A2y}\\
A_3(u)&=36-\frac{18g_2}{u(u-2)}+\frac{9g_2^2}{4u^2(u-2)^2}-\frac{9g_3^2}{16u^2(u-1)^2(u-2)^2}.
 \end{align}
 Note that $A_2(u)$ contains the operator $y$. Therefore, the diagonalization of $A_2(u)$ will provide the eigenvalues of $y$.
 \subsection{Bethe ansatz}
 The idea behind the analytical Bethe ansatz is to first find a particular eigenvalue of the operator to be diagonalized (in our case, by acting on the $su(3)$ highest weight vector), and then to assume that the general formula for the eigenvalue function is a dressed version of the latter. The case for $A_1(u)$ appears as a special case of integrable quantum spin chains with soliton non preserving boundary conditions, which have been examined using the analytical Bethe ansatz in \cite{ACDFR}. Knowing that $A_1(u)$ is already  diagonal in a given $su(3)$ irreducible representation, it would seem of no interest to use the analytical Bethe ansatz at all. However, the exact same dressing functions found for $A_1(u)$ can be used to diagonalize $A_2(u)$, which contains the missing label operator $y$. We note however that this method does not allow the diagonalization of the operator $x$ as it does not appear in the context of such integrable systems. 
 We now present (without proof) the steps in the application of the analytical Bethe ansatz.\\
 
Using equations \eqref{hw1}-\eqref{hw3}, as well as the expression for $A_1(u)$ in terms of $E_{ij}$ given in \eqref{A1Eij}, the action of $A_1(u)$ on the vector of highest weight in an $su(3)$ irrep $(\lambda, \mu)$ reads as:
 \begin{equation}
  A_1(u)V_{\lambda,\mu}=  \Lambda^0_1(u)V_{\lambda,\mu},
 \end{equation}
 where 
 \begin{equation}
     \Lambda^0_1(u)=3 - \frac{  2\alpha_{33}\alpha_{11} + \alpha_{22}^2 + \alpha_{11} - \alpha_{33}}{u^2}
 \end{equation}
 and $\alpha_{ii}$ are defined as \eqref{alphaii}.
 As expected, this eigenvalue is equal to  \eqref{A1su3}, where $g_2$ takes its value \eqref{g2andg3} and $\mathbf{L}^2=(\lambda+\mu)(\lambda+\mu+1)$.
 This eigenvalue can be decomposed as follows:
 \begin{equation}
     \Lambda^0_1(u)=\frac{2u-1}{2u^3}\sigma_1(u)+\frac{\sigma_2(u)}{u^2}+\frac{2u+1}{2u^3}\sigma_3(u)
 \end{equation}
 where
 \begin{eqnarray}
 \sigma_1(u)=(u+\alpha_{11})(u-\alpha_{33}),\qquad \sigma_2(u)=(u+\alpha_{22})(u-\alpha_{22}),\qquad
 \sigma_3(u)= (u+\alpha_{33})(u-\alpha_{11}).
 \end{eqnarray}
 Then the ansatz used in \cite {ACDFR2} is that  the general eigenvalue $ \Lambda_1(u)$ has the following ``dressed'' form
 \begin{equation}
     \Lambda_1(u)=\frac{2u-1}{2u^3}D_1(u)\sigma_1(u)+\frac{D_2(u)\sigma_2(u)}{u^2}+\frac{2u+1}{2u^3}D_3(u)\sigma_3(u),
 \end{equation}
with
 \begin{align}
     D_1(u)&= D_3(-u)=\prod_{j=1}^M\frac{u-u_j-\frac{1}{2}}{u-u_j+\frac{1}{2}},\\
     D_2(u)&=\prod_{j=1}^M\frac{(u-u_i+\frac{3}{2})(u+u_j-\frac{3}{2})}{(u-u_j+\frac{1}{2})(u+u_j-\frac{1}{2})}.
 \end{align}
 It seems that the dressing functions create additional poles. In order for the residues at these poles to vanish, constrains between the  parameters $u_j$ are necessary. These equations are known as Bethe equations and their solutions are called Bethe roots. In our case they read as follows:
\begin{equation} \label{Betheequations}
   \frac{(u_p-\frac{1}{2}-\alpha_{33})(u_p-\frac{1}{2}+\alpha_{11})}{(u_p-\frac{1}{2})^2-\alpha_{22}^2}=\prod_{ \substack{j=1\\j\neq p}}^M\frac{(u_p-u_j+1)(u_p+u_j-2)}{(u_p+u_j-1)(u_p-u_j-1)},\qquad 1\leq p \leq M.
\end{equation}
The physical Bethe roots correspond to the Bethe roots that are pairwise distinct. The different physical Bethe roots will provide the different eigenvalues. The ansatz also states  that the whole spectrum is obtained in this way. \\

The parameter $M$ is related to the eigenvalue of the total angular momentum in a given $so(3)$ representation. Indeed,
one can repeatedly apply L'Hospital rule to compute  the following limit  and get:
\begin{equation}
   \lim_{u \to \infty}(u^2\Lambda_1(u)-3u^2)=g_2-2\mathbf{L}^2,
\end{equation}
where $\mathbf{L}^2=L(L+1)=(\lambda+\mu-M)(\lambda+\mu-M+1)$.\\

We now proceed in a similar fashion to compute the eigenvalue of $A_2(u)$. Acting on the highest weight, we first get:
\begin{equation}
    A_2(u)V_{\lambda,\mu}=\Lambda_2^0 V_{\lambda,\mu},
\end{equation}
where the eigenvalue can be decomposed as follows
\begin{equation}
    \Lambda_2^0=\frac{2(1-2u)}{u^2(u-1)^2}\sigma_1(u)\sigma_3(u-1)-\frac{4}{u^2(u-1)}\sigma_1(u)\sigma_2(u-1)-\frac{4}{u(u-1)^2}\sigma_2(u)\sigma_3(u-1).
\end{equation}
We remark that the same functions $\sigma_i(u)$ appear in the eigenvalues $\Lambda_1^0(u)$ and $\Lambda_2^0(u)$. The ansatz to get all the eigenvalues $\Lambda_1(u)$ consists in dressing each $\sigma_i(u)$ by $D_i(u)$. We then make use of the same dressing functions for $\Lambda_2$, namely:
\begin{align}
    \nonumber \Lambda_2&=\frac{2(1-2u)}{u^2(u-1)^2}D_1(u)\sigma_1(u)D_3(u-1)\sigma_3(u-1)-\frac{4}{u^2(u-1)}D_1(u)\sigma_1(u)D_2(u-1)\sigma_2(u-1)\\
     &-\frac{4}{u(u-1)^2}D_2(u)\sigma_2(u)D_3(u-1)\sigma_3(u-1).
\end{align}
The Bethe roots $u_i$ are solution the same Bethe equations \eqref{Betheequations} as those for $A_1(u)$, insuring that the residues at the poles of the dressing functions $D_i(u)$ vanish. To support this ansatz, let us mention that the same type of dressing appears in the algebraic Bethe ansatz for the Bethe subalgebra of the Yangian studied in \cite{MTV}. 

As explained previously, using equation \eqref{A2y}, we can extract the eigenvalue of the missing label operator $y$ from $\Lambda_2(u)$.
As a check, we solved the Bethe equations for a few low values of $\lambda, \mu$. We wrote them in terms of the elementary symmetric polynomials and solved them using the Gr\"obner basis as explained in \cite{JiangZhang}. One verifies that the obtained eigenvalues of the operator $y$ (see table \ref{tab:BAexamples}) coincide with the ones obtained from direct diagonalization of the tridiagonal matrix $y$ given in \eqref{ymatrixcoeffs1}-\eqref{ymatrixcoeffs2}.\\

\begin{table}[!ht]
\begin{center}
\begingroup
\renewcommand{\arraystretch}{0.8} 
\begin{tabular}{||c | c| c| c | c | c||} 
 \hline
   &&&&&\\[0.5ex] 
$(\lambda,\mu)$ & $(\alpha_{11},\alpha_{22}, \alpha_{33})$ & L &  M  & Elementary symmetric polynomials & Eigenvalues of $y$ \\ &&&&   w.r.t. the Bethe roots $(e_1, e_2, \dots, e_{M})$ &\\
 [0.5ex] 
 \hline\hline &&&&&\\
 (1,0) & ($\frac{2}{3}$, $-\frac{1}{3}$, $-\frac{1}{3})$ & 1 & 0 & - & $\frac{5}{64}$ \\
 &&&&&\\
  \hline   &&&&&\\
  \multirow{3}{*}{(2,1)} &\multirow{3}{*}{($\frac{5}{3}$, $-\frac{1}{3}$, $-\frac{4}{3})$} & 1 & 2 & $\left(\frac{2}{3}, -\frac{13}{36}\right)$ & $\frac{31}{192}$ \\
  &&&&&\\
\cline{3-6}
  &&&&&\\
&& 2 & 1 & $(-\frac{5}{18})$ & $-\frac{35}{64}$\\
  &&&&&\\
\cline{3-6}
 &&&&&\\
&& 3 & 0 & -& $\frac{37}{64}$\\
 &&&&&\\
  \hline   &&&&&\\
  \multirow{4}{*}{(2,2)} & \multirow{4}{*}{(2, 0, $-2$)} & 0&4&($\frac{4}{3}$, $ -\frac{97}{90}$,   $-\frac{11}{5}$, $\frac{ 521}{3600})$&$\frac{29}{64}$\\
  &&&&&\\
\cline{3-6}
  &&&&&\\
  &&\multirow{2}{*}{2}  &\multirow{2}{*}{2}  & $(\frac{2}{3}, -\frac{7}{12})$&$\frac{29}{64}$\\
  &&&&&\\
\cline{5-6}
 &&&&&\\
 & & & &$(-\frac{6}{7}, -\frac{29}{196})$&$-\frac{99}{64}$\\
  &&&&&\\
\cline{3-6}
 &&&&&\\
  & & 3&1&$(-\frac{1}{2})$&$-\frac{75}{64}$\\
  &&&&&\\
\cline{3-6}
 &&&&&\\
  & & 4&0&-&$\frac{69}{64}$\\
  &&&&&\\
 \hline\hline 
\end{tabular}
\endgroup
\end{center}

\caption{\label{tab:BAexamples} Examples of  solutions to Bethe equations \eqref{Betheequations} and the corresponding eigenvalue of $y$.}
\end{table}
\section{Conclusion}
Summing up, we have shown how the analytical Bethe ansatz could be generalized so as to provide a powerful method to diagonalize the $SO(3)$ scalar operator of degree four in the $su(3)$ generators whose eigenvalues give the basis vectors of irreducible modules corresponding to the $SU(3) \supset SO(3) \supset SO(2)$ non-canonical subgroup chain. This adds significant new understanding to an important and thoroughly studied problem by showing how the Bethe ansatz techniques can advance its resolution.

There are other well known and physically important occurrences of missing label problems such as the ones corresponding to the subgroup chains: $SU(4) \supset SU(2) \otimes SU(2)$, $O(5) \supset SU(2) \otimes U(1)$, $O(5) \supset O(3)$, etc. It would be of significant interest to explore further the applicability of the Bethe ansatz methods to the resolution of these and other instances where the degeneracies arise. 

The algebra generated by $x$ and $y$ with relations given in \eqref{eq:algebra} brings reminiscences of algebras of Heun--Askey--Wilson type \cite{BTVZ} and must encode the properties of the overlaps between the two basis that are made out of eigenvectors of $x$ and $y$ respectively. It would be of interest to understand this more deeply.

In this vein, understanding the structure of the algebra associated to the overlaps between the basis corresponding to the canonical chain $SU(3) \supset SU(2) \otimes U(1)$ and the non-canonical one, $SU(3) \supset SO(3)$, is worth attention and would open up many new avenues. We shall keep looking into those questions.

\
\medskip

 \textbf{Acknowledgments:} N.C. is partially supported by Agence National de la Recherche Projet AHA
ANR-18-CE40-0001. D.S.K. benefited from a CGS M scholarship from the Natural Science and Engineering Research Council (NSERC) of Canada during the course of these investigations. L.V. holds a discovery grant from NSERC.

\end{document}